# The Inverse Task of Magnetic Force Microscopy Data

Konstantin Nefedev[a], Vitalii Kapitan[b], Yuriy Shevchenko[c]

Far Eastern Federal University, School of Natural Sciences,

Department of Computer Systems, Russia, 690950, 8 Sukhanova st.

[a]nefedev.kv@dvfu.ru, [b]kapitan.vyu@dvfu.ru, [c]shevchenko.ya@dvfu.ru

**Abstract.**

The computer processing of images of magnetic force microscopy of cobalt nanodots was performed. The solution of reverse task of magnetic force microscopy is obtained for surface points of samples. Superposition of fields, which are generated by a system of magnetic moments in the selected point in space, causes a linear dependence of the force gradient of the dipole-dipole interaction between the components of the vectors.

**Introduction**

Serious changes in the information technologies are related with success researches of magnetic particles [1], and a further magnetic logic development will be associated with magnetic nanostructures. By this reason, both theoretical and experimental research of ensembles and single particles are need, because it is details for the constructing of nano-architectures [2-4]. When the distance between elements is small, the magnitostatic interaction can cause the transform from individual element behavior to collective behavior of the ensemble. The magnetostatic interaction combined with other interaction types is poorly investigated nowadays [5,6]. In this work the attempt of solving of inverse problem of magnetic force microscopy data was made.

**Model**

The model assumes the sample separation (nanoparticle) and the magnetic-force tip partitions on $N_s$ and $N_t$ magnetic dipoles (magnetization vectors), respectively. It is assumed, that each tip discretization element independently magnetostatical interacts with each discretization element of the nanopartical. In the model, magnetic dipoles of the tip have non-zero Z component ($m_{zj}=1$). The system of linear equations (SLE) are formed by following way: the square area 45x45 pixels (Fig. 1b) is forming from source magnetic force microscopy (MFM) image file 89x31 pixels (Fig 1a), then the round area is outlines, which consists of 2025 gradient values (because there is three vector components: $\{m_{xj}, m_{yj}, m_{zj}\}$) (Fig. 1c). Totally, 675 points are considered. Thus, the SLE consist of 2025 equations (675 equations and 3 vector components).

**Magnetic states calculation using MFM data**

It is well-known that impossible to reconstruct the magnetization distribution from both volume and surface charges. But, it is possible to reconstruct magnetization for a 2D system. It has been proved that the 2D task has only one decision [7].

The problem reduced to solving the equations:

$$m_{xj} = \frac{15xz(-3(x^2+y^2)+4z^2)}{\sqrt{(x^2+y^2+z^2)^9}}, \qquad (1)$$

$$m_{yj} = \frac{15yz(-3(x^2+y^2)+4z^2)}{\sqrt{(x^2+y^2+z^2)^9}}, \qquad (2)$$

$$m_{zj} = \frac{3(x^4+3y^4+6x^2(y^2-4z^2)-24y^2z^2+8z^4)}{\sqrt{(x^2+y^2+z^2)^9}}. \qquad (3)$$

The computer software used the equations with given value of force gradient and solved it by modified Gauss method with a choice of the support element. The system of linear equations are created based on (1-3) and data, which taken from a raw image (Fig. 1a).

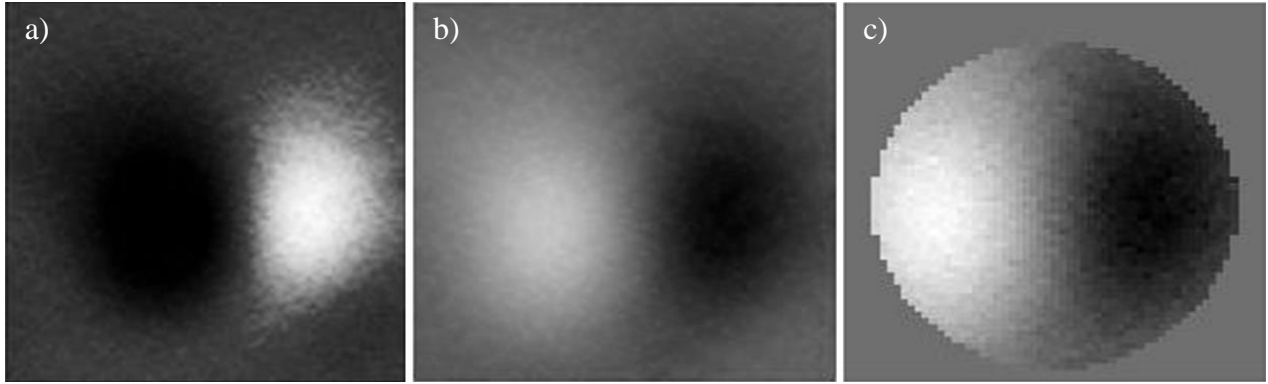

Fig. 1 - The scheme of pixels selections. MFM data used from [8].
a) source image, b) selected squared pixel field, c) rounded pixel field

The modified Gauss method with support element was using to find a root for a SLE. This algorithm allows to solve SLE which contains zeros on the main diagonal. The following algorithm was used in our software: it reads magnitostatic interaction gradient values from the raw data file, forms the circular area which consist of most significant for total magnetization gradients (Fig. 1c), and forms the free members matrix from getting gradient values. Then coefficients of the system are formed through equation (1-3) so that calculated values will be in the circular area similar. SLE was solved by Gauss method. Follow parameters was calculated based on the solution:
  1) Magnetization vector components for various heights of the probe over the nanodot (normalization is performed by one).
  2) Averaged over the nearest neighbor's components of magnetization vectors for different heights over the probe.
Magnetization vectors was calculated and modules of these vectors was visualized, i.e. each vector module equals 1 and only their orientation in space for different heights of the probe over the nanodot was used. For an one-thread software, the execution time was: 4 minutes for 2025 equations SLE, 1 hour for 5184 equations, 6 hours for 11664 equations SLE. Linear code run in one-thread mode at one mainframe cluster's node based on Intel Xenon E5410 2.33GHz, operated by Red Hat Linux.

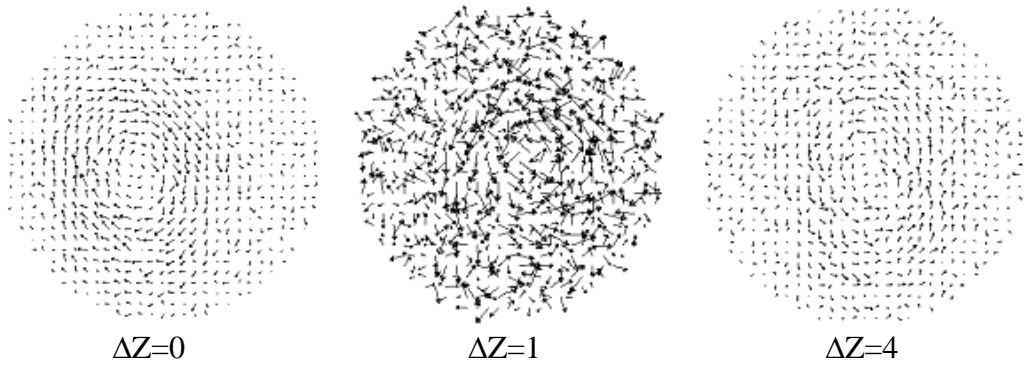
ΔZ=0　　　　　　　　ΔZ=1　　　　　　　　ΔZ=4

Fig. 2 – Magnetization vectors for various heights ΔZ of probe over the nanodot (normalized per unit).

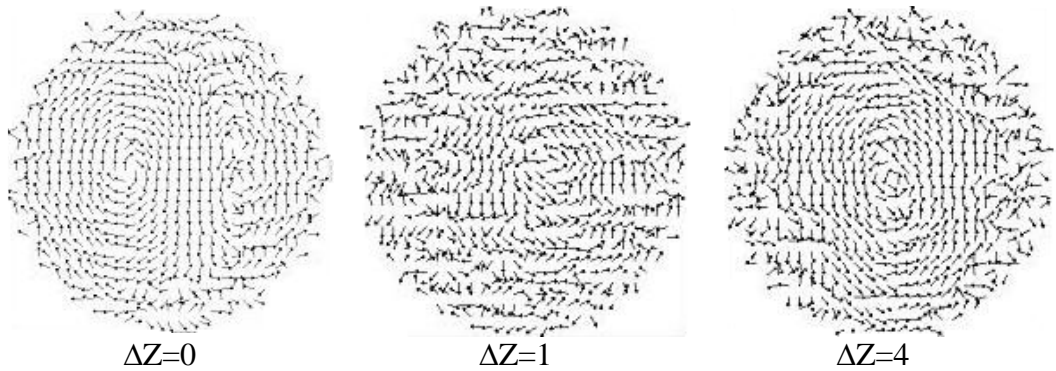
ΔZ=0　　　　　　　　ΔZ=1　　　　　　　　ΔZ=4

Fig. 3 – Averaged by nearest-neighbor magnetization vector components for various heights ΔZ of probe over the nanodot.

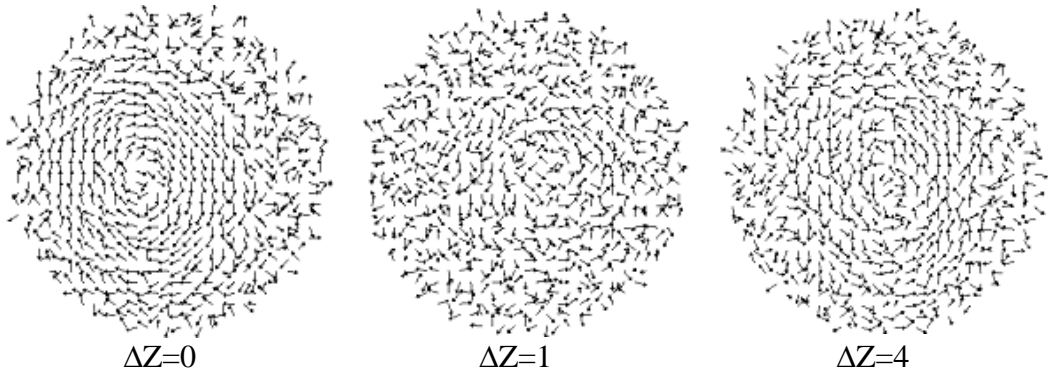
ΔZ=0　　　　　　　　ΔZ=1　　　　　　　　ΔZ=4

Fig. 4 – Magnetization vector modules calculation, i.e. each vector module is 1 and only their orientation in space for different heights of the probe over the nanodot was took in attention.

**Conclusions**

For right experimental data interpretation, it is necessary to solve the reconstruction task of magnetization distribution, using experimental MFM contrast, for difficult magnetic nanodot states case or virtual magnetic experiments with arrays. The magnetic states of individual nanoparticle of cobalt, in classical magnetic dipoles model were calculated. The dependence of the gradient of dipole-dipole interaction power on vector components is due to the field superposition, created by the magnetic moment system at the faced point of space. Considered model demonstrates the spatial MFM-contrast distribution uniqueness for selected magnetic moment configuration, which means uniqueness of magnetic state, i.e. magnetic configuration reconstruction from known experimental MFM-contrast distribution for slim magnetic nanoparticles.